\documentstyle[pre,aps,tighten,preprint]{revtex}
\begin{document}
\draft
\title{Non-perturbative calculation of the probability distribution\\
of plane-wave transmission through a disordered waveguide}
\author{S. A. van Langen, P. W. Brouwer, and C. W. J. Beenakker}
\address{Instituut-Lorentz, University of Leiden, P.O. Box 9506, 2300 RA
Leiden, The Netherlands}
\maketitle

\begin{abstract}
A non-perturbative random-matrix theory is applied to the transmission of a
monochromatic scalar wave through a disordered waveguide. The probability
distributions of the transmittances $T_{mn}$ and $T_n=\sum_m T_{mn}$ of an
incident mode $n$ are calculated in the thick-waveguide limit, for broken
time-reversal symmetry. A crossover occurs from Rayleigh or Gaussian statistics
in the diffusive regime to lognormal statistics in the localized regime. A
qualitatively different crossover occurs if the disordered region is replaced
by a chaotic cavity. \bigskip
\pacs{PACS numbers: 42.25.Bs, 5.45.+b, 72.15.Rn, 78.20.Dj}
\end{abstract}

The statistical properties of transmission through a disordered waveguide have
been extensively studied since 1959, when Gertsenshtein and
Vasil'ev \cite{gertsenshtein}
computed the probability distribution $P(T)$ of the transmittance $T$ of a
single-mode waveguide. It turned out to be remarkably difficult to extend this
result to the $N$-mode case. Instead of a single transmission amplitude $t$
and transmittance $T=|t|^2$, one then has an $N\times N$ transmission matrix
$t_{mn}$ and three types of transmittances
\begin{eqnarray}
&&T_{mn}=|t_{mn}|^2,\ T_n=\sum_{m=1}^N|t_{mn}|^2,\ T=\sum_{n,m=1}^N|t_{mn}|^2.
\end{eqnarray}
All three transmittances have different probability distributions, which can
be measured in different types of experiments: If the waveguide is illuminated
through a diffusor, the ratio of transmitted and incident power equals $T/N$,
because the incident power is equally distributed among all $N$ modes.
(For electrons, $T$ is the conductance in units of $2e^2/h$.) If the incident
power is entirely in mode $n$, then the ratio of transmitted and incident
power equals $T_n$. For $N\gg 1$ this corresponds to illumination by a
plane wave. Finally, $T_{mn}$ measures the speckle pattern (the fraction of
the power incident in mode $n$ which is transmitted into mode $m$).

The complexity of the multi-mode case is due to the strong coupling of the
modes by multiple scattering. While in the single-mode case the localization
length $\xi$
is of the same order of magnitude as the mean free path $l$, the mode coupling
increases $\xi$ by a factor of $N$. If $N\gg 1$, a waveguide of length $L$
can be in two distinct regimes: the diffusive regime $l\ll L\ll Nl$ and the
localized regime $L\gg Nl$. The average of each of the three transmittances
decays linearly with $L$ in the diffusive regime and exponentially in the
localized regime. In an important development, Nieuwenhuizen and
Van Rossum \cite{nieuwenhuizen} (and more recently Kogan and
Kaveh \cite{kogan}) succeeded in computing the probability distributions
$P(T_{mn})$ and $P(T_{n})$ for plane-wave illumination in the diffusive
regime. The former is exponential (Rayleigh's law) with non-exponential
tails, while the latter is Gaussian with non-Gaussian tails. The existence
of such anomalous tails has been observed in optical
experiments \cite{genack,deboer} and in numerical simulations \cite{edrei}.
{}From the simulations, one expects a crossover to a lognormal distribution
on entering the localized regime. Since the theory of
Refs.\ \cite{nieuwenhuizen,kogan} is based on a perturbation expansion in the
small parameter $L/Nl$, it cannot describe this crossover which occurs
when $L/Nl\simeq 1$.

It is the purpose of the present paper to provide a non-perturbative
calculation
of $P(T_{mn})$ and $P(T_n)$, which is valid all the way from the diffusive
into the localized regime, and which shows how the Rayleigh and
Gaussian distributions of $T_{mn}$ and $T_n$ evolve into the same lognormal
distribution as $L$ increases beyond the localization length $\xi\simeq Nl$.
We expect that $P(T)$ also evolves from a Gaussian to a lognormal
distribution, but our calculation applies only to the plane-wave transmittances
$T_{mn}$ and $T_n$, and not to the transmittance $T$ for diffuse illumination.
For technical reasons, we need to assume that time-reversal symmetry is broken
by some magneto-optical effect. Similar results are expected
in the presence of time-reversal symmetry, but then a non-perturbative
calculation becomes much more involved. We make essential use of the
quasi-one-dimensionality of the waveguide (length $L$ much greater than width
$W$) and assume weak disorder (mean free path $l$ much greater than wavelength
$\lambda$). The localization which occurs in unbounded media when
$l\lesssim\lambda$ requires a very different non-perturbative
approach \cite{altshuler}.

A related problem of experimental interest is the transmittance of a cavity
coupled to two $N$-mode waveguides without disorder. If the cavity has an
irregular shape, it has a complicated ``chaotic'' spectrum of eigenmodes.
At the end of the paper we compute
$P(T_{mn})$ and $P(T_n)$ for such a chaotic cavity and contrast the
results with the disordered waveguide, which we consider first.

Our calculation applies results from random-matrix theory for the statistics
of the transmission matrix. This matrix $t=u\sqrt{\tau}v$ is the product of two
unitary matrices $u$ and $v$, and a matrix
$\tau={\rm diag}(\tau_1,\tau_2,\ldots\tau_N)$ containing the transmission
eigenvalues.
It describes the transmission of electrons or electromagnetic radiation, to
the extent
that the effects of electron-electron interaction or polarization can be
disregarded. The two plane-wave transmittances which we consider are
\begin{eqnarray}
&&T_{mn}^{\vphantom{\ast}}=\sum_{k,l}u_{mk}^{\vphantom{\ast}}u_{ml}^\ast
v_{kn}^{\vphantom{\ast}} v_{ln}^\ast
\sqrt{\tau_k^{\vphantom{\ast}}\tau_l^{\vphantom{\ast}}},~~
T_n^{\vphantom{\ast}}= \sum_k
|v_{kn}^{\vphantom{\ast}}|^2 \tau_k^{\vphantom{\ast}}.
\label{tmn}
\end{eqnarray}
We seek the probability distributions
\begin{mathletters}
\begin{eqnarray}
&&P({\cal T}_{mn}) = \langle\delta ({\cal T}_{mn}-N^2T_{mn})\rangle ,\\
&&P({\cal T}_n) = \langle\delta ({\cal T}_n - NT_n)\rangle ,
\end{eqnarray}
\end{mathletters}%
of the normalized transmittances ${\cal T}_{mn}=N^2T_{mn}$ and
${\cal T}_n=NT_n$. (These conventions differ by a factor $l/L$ with
Refs.~\cite{nieuwenhuizen,kogan}.) The brackets $\langle\cdots\rangle$ denote
an average over the disorder.
In the quasi-one-dimensional limit of a waveguide which
is much longer than wide, the matrices $u$ and $v$ are uniformly distributed
in the unitary group \cite{mello}. The joint probability distribution of
the transmission eigenvalues evolves with increasing
$L$ according to the Dorokhov-Mello-Pereyra-Kumar (DMPK) equation \cite{dmpk}.
The average can be performed in two steps, first over $u$ and $v$, and
then over the transmission eigenvalues $\tau_k$.

The first step was done by Kogan and Kaveh \cite{kogan}. The result
is an expression for the Laplace transform of $P({\cal T}_n)$,
\begin{equation}
F(s) = \int_0^\infty\! d{\cal T}_n\,\exp\left(-s{\cal T}_n\right) P({\cal
T}_n),
\label{laplace}
\end{equation}
which in the thick-waveguide limit
($N\rightarrow\infty, L/l\rightarrow\infty$, fixed $Nl/L$) is exactly
given by
\begin{equation}
F(s) = \left\langle {\textstyle\prod_k} (1+s\tau_k)^{-1} \right\rangle.
\label{generating}
\end{equation}
The same function $F(s)$ also determines $P({\cal T}_{mn})$, which in the
same limit is related to $P({\cal T}_n)$ by \cite{kogan}
\begin{equation}
P({\cal T}_{mn}) = \int_0^\infty d{\cal T}_n^{\vphantom{-1}}\, {\cal T}_n^{-1}
\exp \left( -{\cal T}_{mn}^{\vphantom{-1}}/{\cal T}_n^{\vphantom{-1}}
 \right) P({\cal T}_n^{\vphantom{-1}}).
\label{mapping}
\end{equation}

The next step, which is the most difficult one, is to average
over the transmission eigenvalues in Eq.\ (\ref{generating}).
The result depends on whether time-reversal symmetry is present or
not (indicated by $\beta =1$ or $2$, respectively). In
Refs.\ \cite{nieuwenhuizen,kogan}, $\ln F$ was evaluated to leading order in
$L/Nl$, under the assumption that the waveguide length $L$ is much less than
the localization length $\xi\simeq Nl$. Here we relax this assumption.

We consider the case of broken time-reversal symmetry ($\beta=2$).
Then the probability distribution of the $\tau_k$'s is known exactly, in
the form of a determinant of Legendre functions $P_\nu$ \cite{beenakker1}.
Still, to compute expectation values
with this distribution is in general a
formidable problem. It is a lucky coincidence that the
average (\ref{generating}) which we need can be evaluated exactly.
This was shown by Rejaei \cite{rejaei},
using a field-theoretic approach which leads to a supersymmetric non-linear
$\sigma$ model \cite{efetov}.
It was recently proven \cite{brouwer}
that this supersymmetric theory is equivalent to the DMPK-equation used in
Ref.\ \cite{beenakker1}.
{}From Rejaei's general expressions we find
\begin{eqnarray}
&&F(s)=1\!-\!2s\!\sum_{p=0}^\infty\int_0^\infty\!\!\!\!
dk\ f_p(k)\tanh ({\case{1}{2}\pi k}) P_{\frac{1}{2}({\rm i}k-1)}(1\!+\!2s) ,
\nonumber \\
&&f_p(k)=\frac{(2p+1)k}{(2p+1)^2+k^2}
\exp\left(-\frac{L\left[(2p+1)^2+k^2\right]}{4Nl}\right). \nonumber
\end{eqnarray}
Inversion of the Laplace transform (\ref{laplace}) yields $P({\cal T}_n)$,
\begin{equation}
P({\cal T}_n) =
\sum_{p=0}^\infty\int_0^\infty\!\!\! dk\, f_p(k)\sinh(\case{1}{2}\pi k)
\frac{\partial}{\partial{\cal T}_n}
\frac{2K_{\frac{1}{2}{\rm i}k}(\frac{1}{2}{\cal T}_n)}{(\pi^3 {\cal T}_n {\rm
e}^{{\cal T}_n})^{1/2}},
\label{exact}
\end{equation}
where $K_\nu$ is the Macdonald function. One further integration gives
$P({\cal T}_{mn})$, in view of Eq.\ (\ref{mapping}). Results are plotted in
Fig.\ \ref{fig1}.
The large ${\cal T}_n$ and ${\cal T}_{mn}$ tails are
\begin{mathletters}
\begin{eqnarray}
&&P({\cal T}_{mn})\propto {\cal T}_{mn}^{-3/4}
{\rm e}^{-2\sqrt{{\cal T}_{mn}}},~~~~~{\cal T}_{mn}\gg 1,(Nl/L)^2,
\label{mntail} \\
&&
\lefteqn{
P({\cal T}_n)\propto {\cal T}_n^{-1}{\rm e}^{-{\cal T}_n}, }
\hphantom{P({\cal T}_{mn})\propto {\cal T}_{mn}^{-3/4}
{\rm e}^{-2\sqrt{{\cal T}_{mn}}},~~~~~}
{\cal T}_n \gg 1,Nl/L. \label{ntail}
\end{eqnarray}
\end{mathletters}%
It is worth noting that Fyodorov and Mirlin \cite{fyodorov} found the same tail
as Eq.\ (\ref{mntail}) for the distribution of the local density of electronic
states in a closed disordered wire. It is not clear to us whether this
coincidence is accidental.

The diffusive and localized limits can be computed from
Eq.\ (\ref{generating}) by using the known asymptotic form
of the distribution of the $\tau_k$'s. In contrast to the full
result (\ref{exact}),
which holds for $\beta=2$ only, the following asymptotic expressions hold for
any $\beta$. In the diffusive regime, for $L\ll Nl$, we may expand $\ln F$
in cumulants of the linear statistic $A=\sum_k \ln(1+s\tau_k)$:
\begin{eqnarray}
&&\ln F(s)\equiv\ln\left\langle{\rm e}^{-A}\right\rangle =
-\left\langle A\right\rangle + \case{1}{2}{\rm Var}\ A + {\cal O}(L/Nl).
\label{expansion}
\end{eqnarray}
The mean and variance of $A$ can be computed from the general formulas of
Refs.\ \cite{beenakker1,beenakker2,chalker}:
\begin{mathletters}
\label{meanvariance}
\begin{eqnarray}
&&\left\langle A\right\rangle = \frac{Nl}{L}\,{\rm asinh}^2\sqrt{s}
+ \frac{2-\beta}{4\beta}\!
\ln\left[\frac {{\rm asinh}^2\sqrt{s}} {s(1+s)} \right],\\
&&{\rm Var}\ A = -\frac{1}{\beta}\left[ \ln(1+s) +
 6\ln\left(\!\frac{{\rm asinh}\sqrt{s}}{\sqrt{s}}\right)\right],
\end{eqnarray}
\end{mathletters}%
valid up to corrections of order $L/Nl$. To leading order in $L/Nl$ one has
the $\beta$-independent result of Refs.\ \cite{nieuwenhuizen,kogan},
yielding Gaussian and Rayleigh statistics for $L/Nl\rightarrow 0$.
The $\beta$-dependent terms in Eqs.\ (\ref{meanvariance}) are
the first corrections due to localization effects.
In Fig.\ \ref{fig2} we plot $P({\cal T}_n)$ resulting
from Eqs.\ (\ref{expansion}) and (\ref{meanvariance}).
The $\beta$-independent result of Refs.\ \cite{nieuwenhuizen,kogan}
(not shown) is very close to the $\beta=2$ curve.
This figure indicates that the $\beta$-dependence is essentially quantitative
rather than qualitative.

In the opposite, localized regime ($L\gg Nl$), only a single transmission
eigenvalue contributes significantly to Eq.\ (\ref{generating}). This largest
eigenvalue $\tau$ has the lognormal distribution \cite{stone}
\begin{equation}
P(\ln\tau)= \left(\case{\beta Nl}{8\pi L}\right)^{1/2}
\exp\left[-\case{\beta Nl}{8L}\left(\case{2L}{\beta Nl}+
\ln\tau\right)^2\right].
\label{lognormal}
\end{equation}
It follows that $\ln{\cal T}_{mn}$ and $\ln{\cal T}_n$ are also distributed
according to Eq.\ (\ref{lognormal}) in the localized regime. The approach
to a common lognormal distribution as $L/Nl$ increases is illustrated in
Fig.\ \ref{fig3}, using the exact $\beta=2$ result of Eq.\ (\ref{exact}).

We contrast these results for a disordered waveguide with those for a
chaotic cavity, attached to two $N$-mode leads without disorder.
Following Ref.\ \cite{baranger} we assume that
the $2N\times 2N$ scattering matrix of the cavity is distributed uniformly
in the unitary group if $\beta=2$ or in the subset of unitary and symmetric
matrices if $\beta=1$. Then $P(T_{mn})$ and $P(T_n)$ follow
from general formulas \cite{pereyra} for the distribution of matrix
elements in these socalled ``circular'' ensembles. For $\beta=2$ the result is
\begin{mathletters}
\begin{eqnarray}
&&P(T_{mn})=(2N-1)\left(1-T_{mn}\right)^{2N-2} \label{mncavity}, \\
&&P(T_n)= \case{1}{2}N{2N \choose N}\left[ T_n (1-T_n)\right]^{N-1}.
\label{ncavity}
\end{eqnarray}
\end{mathletters}%
For $\beta=1$ Eq.\ (\ref{mncavity}) should be multiplied by
$\frac{1}{2}F(N-\frac{1}{2},1;2N-1;1-T_{mn})$ and Eq.\ (\ref{ncavity}) by
$\frac{1}{2}F(N-\frac{1}{2},1;N;1-T_n)$, where $F$ is the
hypergeometric function. These are exact results for any $N$.
If $N\rightarrow\infty$, $P(T_{mn})$ is an exponential distribution
with mean $1/2N$, and $P(T_n)$ is a Gaussian with mean $1/2$ and variance
$1/8N$. This is similar to the disordered waveguide, with $N$ playing the
role of $Nl/L$. As shown in Fig.\ \ref{fig4}, the distributions for $N$
of order unity are quite different from those in a disordered waveguide with
$Nl/L$ of order unity. For $N=1$ the distinction between $T_{mn}, T_n$, and $T$
disappears and we recover the results of Ref.\ \cite{baranger}.

In conclusion, we have presented a non-perturbative calculation of the
distributions of the plane-wave transmittances $T_{mn}$ and $T_n$ through
a disordered wave\-guide without time-reversal symmetry, which shows how
the distributions cross over from Rayleigh and Gaussian statistics in the
diffusive regime, to a common lognormal distribution in the localized regime.
Qualitatively different distributions are obtained if the disordered region
is replaced by a chaotic cavity. Existing experiments have been mainly in
the regime $L\ll Nl$ where the perturbation theory of
Refs.\ \cite{nieuwenhuizen,kogan} applies. If the absorption of light in the
waveguide can be reduced sufficiently, it should be possible to enter the
regime $L\simeq Nl$ where perturbation theory breaks down and the crossover
to lognormal statistics is expected.

This research was supported by the ``Ne\-der\-land\-se or\-ga\-ni\-sa\-tie voor
We\-ten\-schap\-pe\-lijk On\-der\-zoek'' (NWO) and by the ``Stich\-ting voor
Fun\-da\-men\-teel On\-der\-zoek der Ma\-te\-rie'' (FOM).

\begin{figure}
\caption{\label{fig1}
Distributions of (a) ${\cal T}_n\equiv NT_n$ for
$L/Nl=$ 0.05, 0.1,  0.5, 1.5, 2.0, and 2.5, and (b)
${\cal T}_{mn}\equiv N^2T_{mn}$
for $L/Nl=$ 0.05, 0.5, 2.5, 5, and 10. Computed from the exact $\beta=2$
expressions (\ref{mapping}) and (\ref{exact}). The dotted curves are the
limits $L/Nl\rightarrow 0$ of an infinitely narrow Gaussian in (a) and
an exponential distribution in (b) (note the logarithmic scale). The inset
in (b) shows the waveguide geometry considered (disordered region is shaded).}
\end{figure}
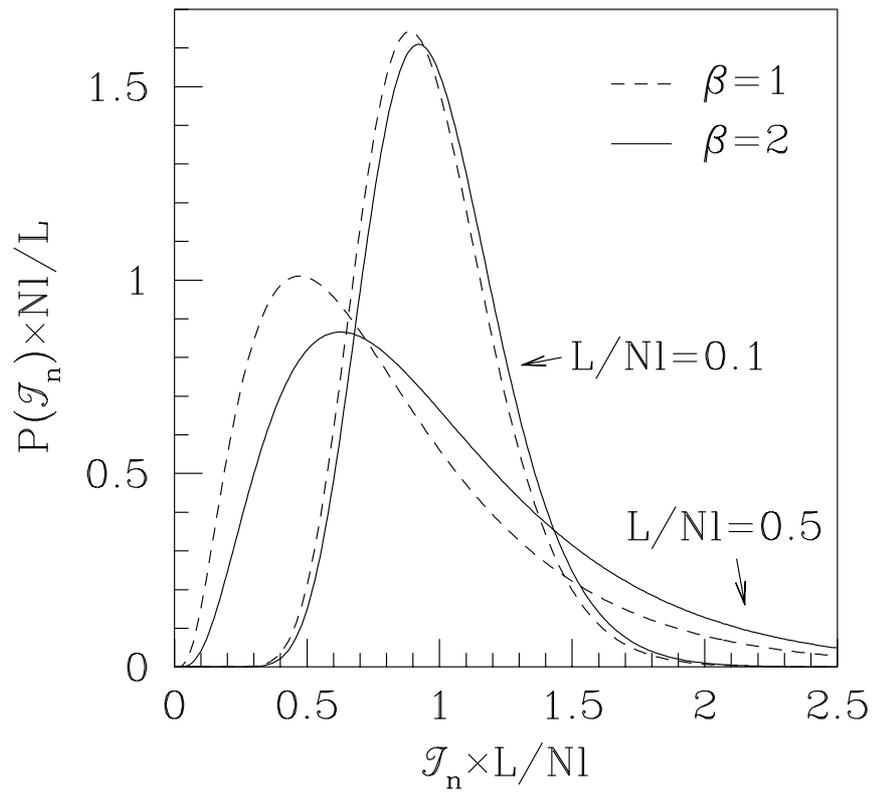
\begin{figure}
\caption{\label{fig2} Distribution of ${\cal T}_n$ calculated from the
perturbation
expansion (\ref{expansion}), (\ref{meanvariance}), for $\beta=1,2$ and
$L/Nl=0.1, 0.5$. }
\end{figure}
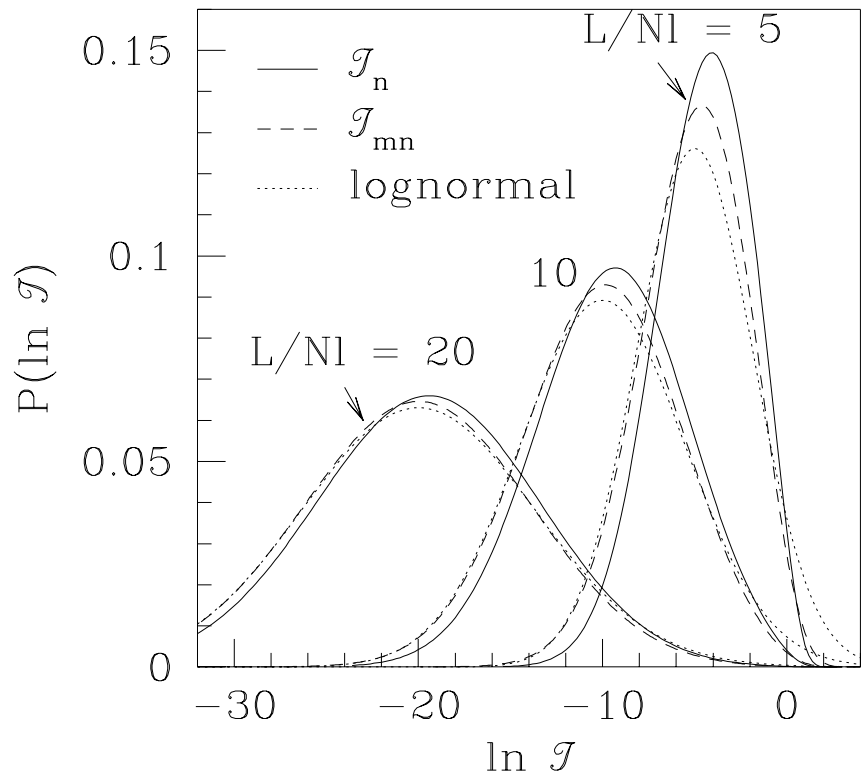
\begin{figure}
\caption{\label{fig3}
Distributions of ${\cal T}_n$ and ${\cal T}_{mn}$ for
$\beta=2$ and $L/Nl=5, 10, 20$, computed from Eqs.\ (\ref{mapping}) and
(\ref{exact}). The dotted curve is the lognormal
distribution (\ref{lognormal}) which is approached as
$L/Nl\rightarrow\infty$.}
\end{figure}
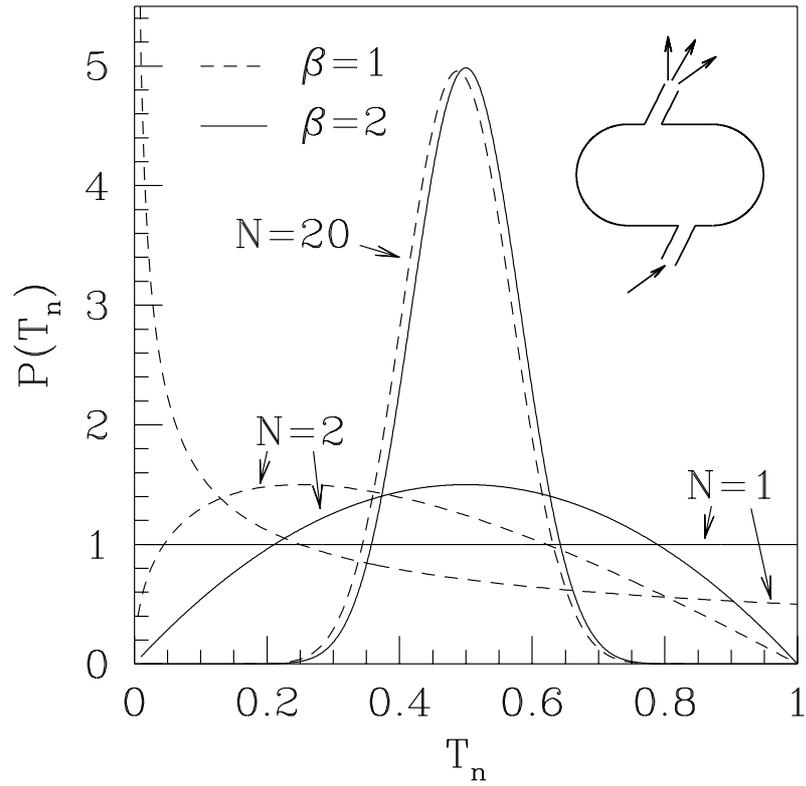
\begin{figure}
\caption{\label{fig4}
Distribution of $T_n$ for a chaotic cavity attached to
two $N$-mode leads (inset). The curves are computed from Eq.\ (\ref{ncavity}),
for $\beta=$ 1, 2 and $N=$ 1, 2, 20.}
\end{figure}
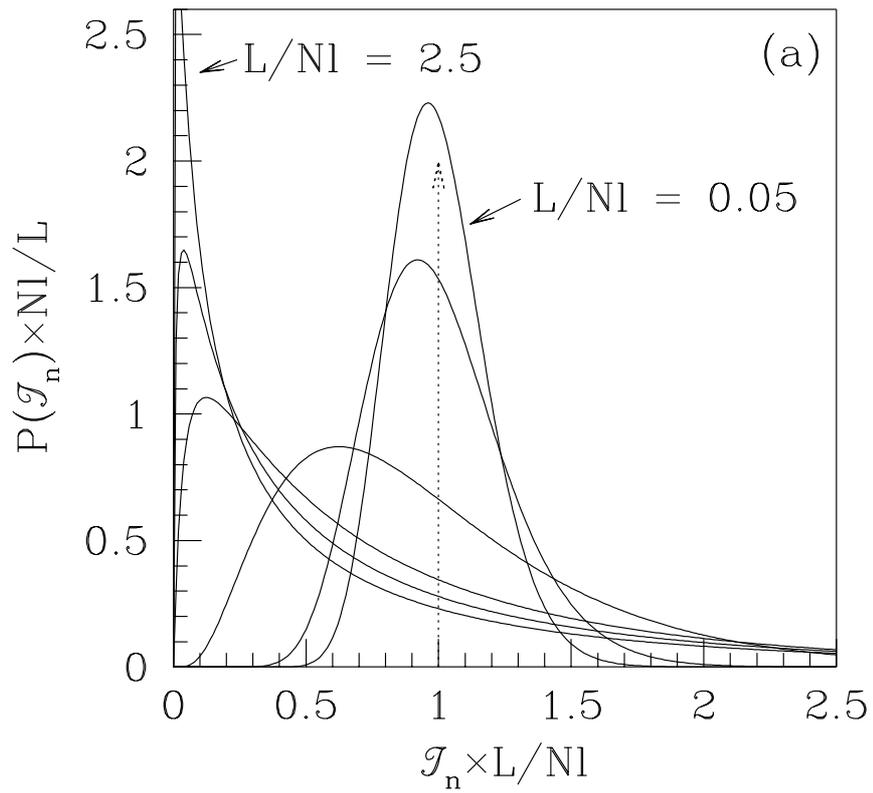
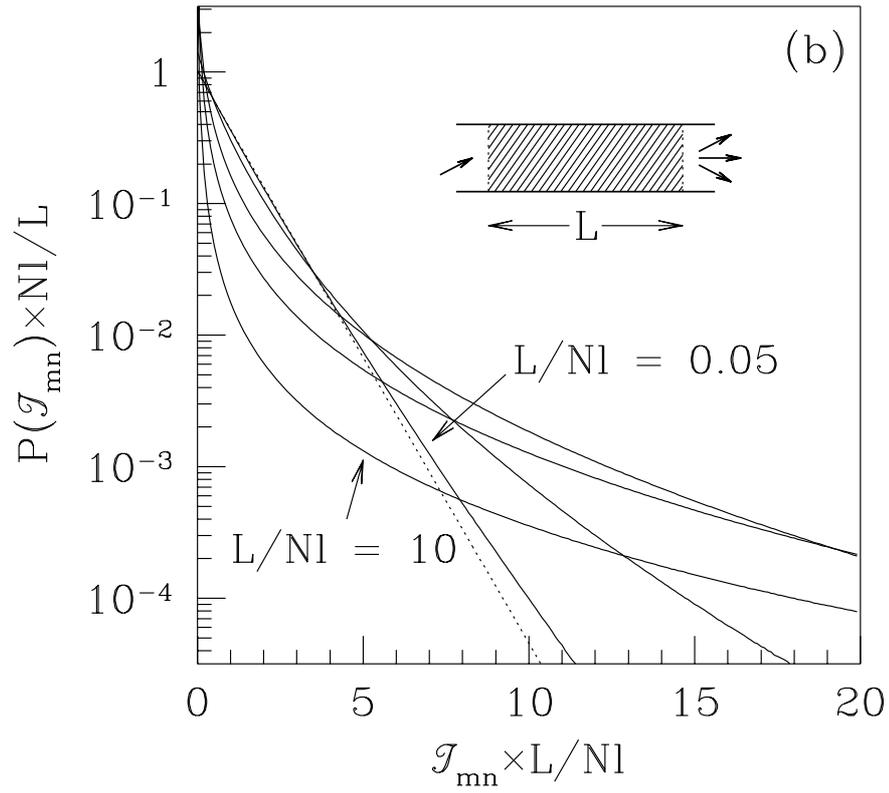
\end{document}